\documentclass[%
 aip,
 amsmath,amssymb,
 reprint,%
]{revtex4-1}

\usepackage{graphicx}
\usepackage{dcolumn}
\usepackage{bm}

\usepackage[utf8]{inputenc}
\usepackage[T1]{fontenc}
\usepackage{mathptmx}

\begin{document}

\preprint{AIP/123-QED}

\title{Pulsed thermal deposition of binary and ternary transition metal dichalcogenide
monolayers and heterostructures}

\author{Niklas Mutz}
\affiliation{Institut f\"{u}r Physik, Institut f\"{u}r Chemie and IRIS Adlershof, Humboldt-Universit\"{a}t zu Berlin, Berlin, Germany}

\author{Tino Meisel}
\affiliation{Institut f\"{u}r Physik, Humboldt-Universit\"{a}t zu Berlin, Berlin, Germany}

\author{Holm Kirmse}
\affiliation{Institut f\"{u}r Physik, Humboldt-Universit\"{a}t zu Berlin, Berlin, Germany}

\author{Soohyung Park}
\affiliation{Institut f\"{u}r Physik, Humboldt-Universit\"{a}t zu Berlin, Berlin, Germany}

\author{Nikolai Severin}
\affiliation{Institut f\"{u}r Physik and IRIS Adlershof, Humboldt-Universit\"{a}t zu Berlin, Berlin, Germany}

\author{J\"{u}rgen P. Rabe}
\affiliation{Institut f\"{u}r Physik and IRIS Adlershof, Humboldt-Universit\"{a}t zu Berlin, Berlin, Germany}

\author{Emil List-Kratochvil}
\affiliation{Institut f\"{u}r Physik, Institut f\"{u}r Chemie and IRIS Adlershof, Humboldt-Universit\"{a}t zu Berlin, Berlin, Germany}

\author{Norbert Koch}
\affiliation{Institut f\"{u}r Physik and IRIS Adlershof, Humboldt-Universit\"{a}t zu Berlin, Berlin, Germany}

\author{Christoph Koch}
\affiliation{Institut f\"{u}r Physik and IRIS Adlershof, Humboldt-Universit\"{a}t zu Berlin, Berlin, Germany}

\author{Sylke Blumstengel}
\email[]{sylke.blumstengel@physik.hu-berlin.de}
\affiliation{Institut f\"{u}r Physik, Institut f\"{u}r Chemie and IRIS Adlershof, Humboldt-Universit\"{a}t zu Berlin, Berlin, Germany}

\author{Sergey Sadofev}
\email[]{sergey.sadofev@physik.hu-berlin.de}
\affiliation{Institut f\"{u}r Physik, Humboldt-Universit\"{a}t zu Berlin, Berlin, Germany}


\begin{abstract}
Application of transition metal dichalcogenides (TMDC) in optoelectronic, photonic or valleytronic devices requires the growth of continuous monolayers, heterostructures and alloys of different materials in a single process. We present a facile pulsed thermal deposition method which provides precise control over the number of layers and the composition of two-dimensional systems. The versatility of the method is demonstrated on ternary monolayers of Mo$_{1-x}$W$_{x}$S$_{2}$ and on heterostructures combining metallic TaS$_{2}$ and semiconducting MoS$_{2}$ layers. The fabricated ternary monolayers cover the entire composition range of $x$ = 0...1 without phase separation. Band gap engineering and control over the spin-orbit coupling strength is demonstrated by absorption and photoluminescence spectroscopy. Vertical heterostructures are grown without intermixing. The formation of clean and atomically abrupt interfaces is evidenced by high-resolution transmission electron microscopy. Since both the metal components as well as the chalcogens are thermally evaporated complex alloys and heterostructures can thus be prepared.
\end{abstract}

\maketitle

Transition metal dichalcogenides (TMDC) span a conductivity range from metal (group V TMDCs) to semiconductor (group VI TMDCs). The optical response of the semiconducting TMDCs is governed up to room temperature by Coulombically bound neutral and charged excitons (trions). They are responsible for the large light-matter interaction cross section and the efficient photoluminescence (PL) in the monolayer regime. The favourable optical properties combine with fairly high charge carrier mobilities up to a few hundred cm$^2$/Vs and the feasibility of $p$- and $n$-doping. \cite{Ye2017, Zhang2018} Furthermore, the lack of inversion symmetry and the thus strong spin-orbit band splitting are responsible for the unique coupling of the spin and valley degrees of freedom.\cite{Cao2012} All these properties combined make TMDCs attractive as ultra-thin, flexible and semitransparent platforms for novel optoelectronic, photonic and valleytronic devices. Like in traditional semiconductor devices, the functionality is achieved by alloy engineering and the combination of different materials in heterostructures. Alloying allows tuning the band structure to achieve the desired optical and electronic properties.\cite{Xi2014, Chen2013} The spin-orbit coupling (SOC) represents a new degree of freedom in TMDCs. Its strength, being dependent mainly on the transition metal $d$-orbitals, is tunable in a wide range (a few hundreds of meV) in ternary Mo$_{1-x}$W$_{x}$X$_{2}$ alloys with X = S, Se.\cite{Wang2015} Thus, the population of bright and dark excitons and the valley polarisation can be tuned. Stacks of different materials are the essential building blocks of all above mentioned devices. Also in this respect, monolayer TMDCs promise a richer behavior than traditional semiconductors because novel electronic properties not found in the constituent layers have been predicted to arise in TMDC heterostructures.\cite{Terrones2013} Furthermore, the combination of semiconducting and metallic TMDCs opens up the opportunity for achieving all-TMDC-based devices. 

So far, TMDC alloys have been synthesized either by mixing different chalcogen atoms (MX1$_{2(1-x)}$X2$_{2x}$) or transition metals (M1$_{1-x}$M2$_{x}$X$_{2}$) using chemical vapour deposition (CVD)\cite{Gong2014, Zhang2015, Wang2016}, low-pressure vapour transport\cite{Feng2014} or pulsed laser deposition\cite{Yao2016} techniques to yield materials in monolayer or bulk form. Vertical TMDC heterostructures have been achieved by mechanical stacking of individual layers\cite{Kang2017} or by direct growth employing CVD\cite{Gong2014a}, metal-organic chemical vapour deposition\cite{Lin2015} and molecular beam epitaxy (MBE)\cite{Vishwanath2016}. Metallic group V TaS$_{2}$ has attracted interest as charge density wave material and CVD growth of mono- and multilayer flakes has been reported.\cite{Fu2016}

\begin{figure}[t]
\centerline{\includegraphics{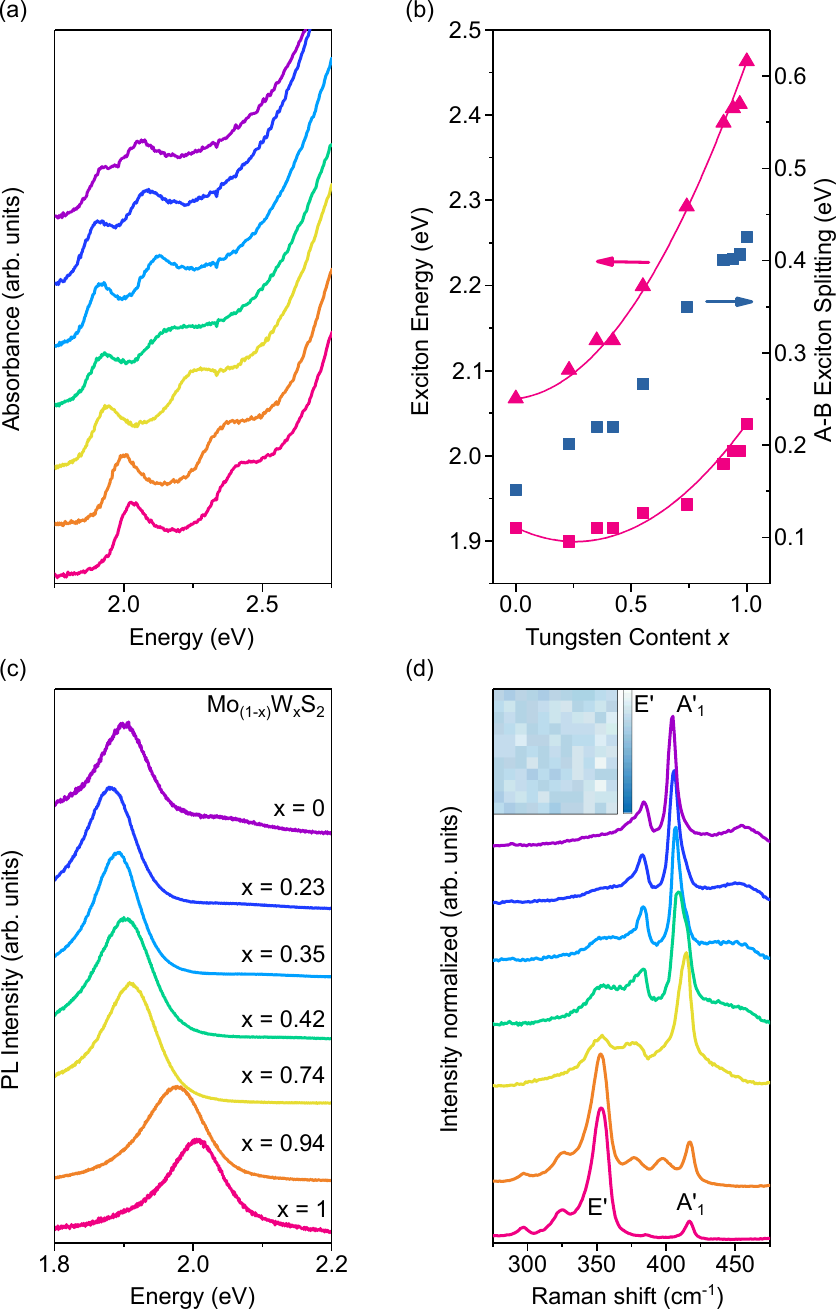}}
\caption{\label{fig:Raman} a) Absorption spectra of Mo$_{1-x}$W$_{x}$S$_{2}$ monolayers with varying composition $x$. b) Transition energies of A and B excitons extracted from absorption measurements and A-B exciton splitting. The solid lines are fits to extract the bowing parameters of the exciton energies. c) PL spectra of the monolayers excited at 2.82 eV. d) Raman spectra acquired with a 2.33 eV excitation laser. The inset shows a Raman map of the intensity ratio of the MoS$_{2}$-like and the WS$_{2}$-like $E'$ peak for a monolayer with $x$ = 0.55. The scan area is 100 $\mu$m $\times$ 100 $\mu$m and the scale bar spans an intensity ratio range from 1.10 to 1.15. The composition of the monolayers according to XPS data is given in (c). The color code is identical in (a), (c) and (d).}
\end{figure}

To pave the way for practical applications of TMDCs in devices, a fabrication method is required which allows preparation of alloys with defined composition as well as continuous growth of multiple layers of alternating materials with clean and atomically sharp interfaces in a single process. In this letter, we report a versatile and easy to implement pulsed thermal deposition (PTD) method for the fabrication of large-area ternary Mo$_{1-x}$W$_{x}$S$_{2}$ monolayers as well as heterostructures involving metallic TaS$_{2}$. The high-vacuum-based process avoiding ambient exposure guaranties minimal interface contamination. Furthermore, the film thicknesses and the compositions can be precisely controlled. By combination of optical and photoemission spectroscopies we show that ternary monolayers can be grown over the entire composition range $x =$ 0...1 without phase separation. High-resolution transmission electron microscopy (HRTEM) reveals moreover atomically sharp interfaces of the heterostructures combining different TMDCs. 

PTD is performed using pulsed direct resistive heating and sublimation of metal elements (Mo, W, Ta), while the group VI element (S) is co-evaporated from a standard Knudsen cell. The metal evaporators are serpentine wire filaments with wire diameters of 1 mm. The distance between the evaporators and the substrate is about 10 cm. The base pressure in the growth reactor is 10$^{-6}$ Torr. The layers are deposited on fused silica substrates kept at a temperature of ca. 400 $^\circ$C during the deposition. In order to precisely control layer thickness and composition, the sublimation of the metals is performed in a pulsed mode. In this way, defined deposition of submonolayers up to several monolayers of TMDCs can be achieved by adjusting the pulse number and duration. Typical deposition parameters for the growth of a monolayer are 4 pulses of a duration of 90 s and a time delay between the pulses of 180 s. The mixed crystals can be either grown in an alternating mode by switching repetitively between sublimation of Mo and W or by simultaneous deposition of both metals. Both approaches yield similar results. 

The composition of the ternary Mo$_{1-x}$W$_{x}$S$_{2}$ layers is determined by X-ray photoemission spectroscopy (XPS). The measurements are performed with a JEOL JPS-0930 system using a monochromatized aluminum K$\alpha$ X-ray source. The peak areas of the Mo 3$p$, W 4$f$, and S 2$p$ core levels corrected with atomic sensitivity factors provided by the JEOL system have been used for estimation of $x$. Since the conductivity of the monolayers is very low, a rigid shift and a broadening of all core level peaks were observed due to sample charging. This leads to an error of up to $\pm$ 5 \% in the estimation of the composition.

\begin{figure*} [t]
\centerline{\includegraphics{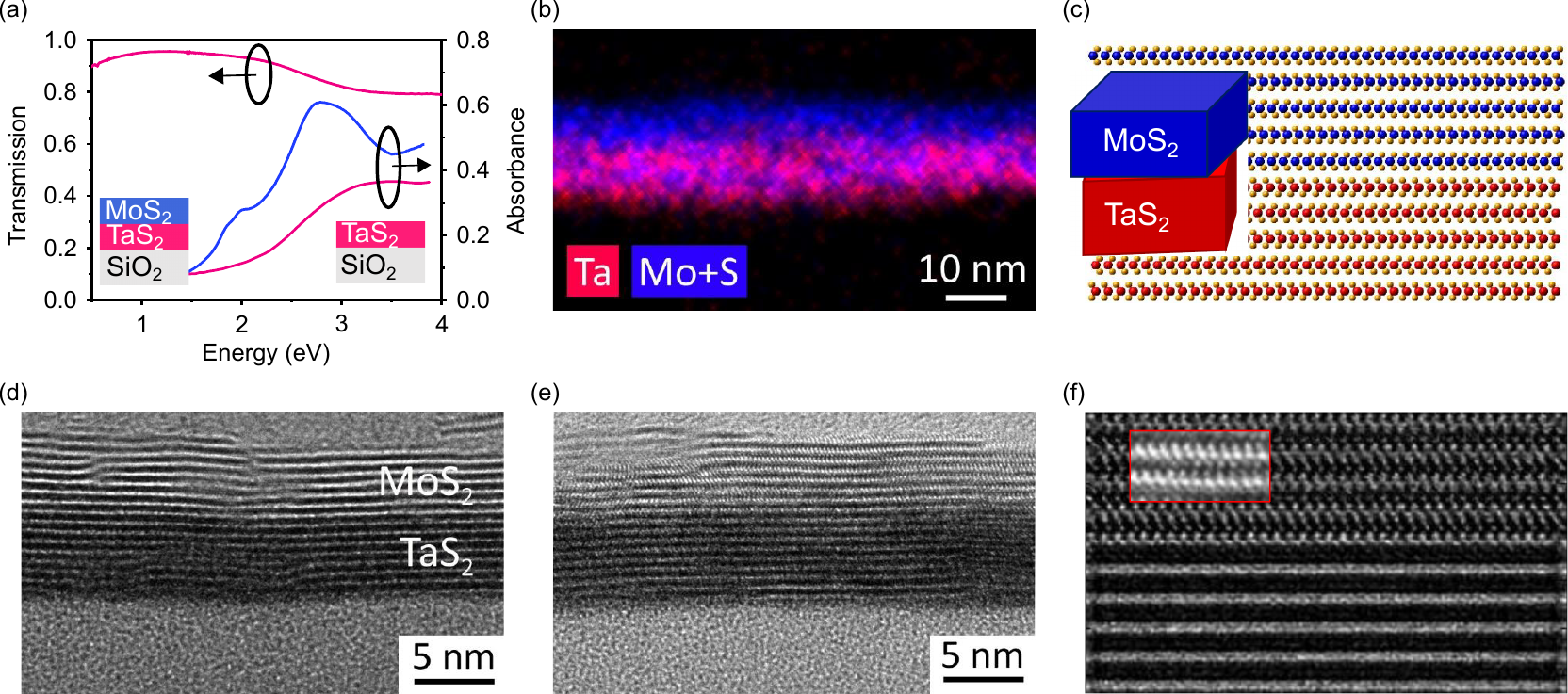}}
\caption{\label{fig:Tantal} a) Transmission spectrum of a TaS$_{2}$ film consisting of 5 monolayers. Absorbance spectrum of a  MoS$_{2}$/TaS$_{2}$ heterostructure in comparison to a spectrum of a sole TaS$_{2}$ film. b) EDXS map of the MoS$_{2}$/TaS$_{2}$ heterostructure. Depicted is the distribution of tantalum (red) and of sulfur and molybdenum (blue). Due to signal overlap, the contributions of the two elements can not be separated. c) Schematic depiction of the crystalline structure of the MoS$_{2}$/TaS$_{2}$ heterostructure. The cartoon in the inset shows the lattice twist between the layers. d, e) Cross-sectional HRTEM images of the heterostructure recorded at two different positions. f) Image contrast simulations of the HRTEM of the heterostructure taking into account a twist between the MoS$_{2}$ and TaS$_{2}$ lattice of 20$^\circ$ as schematically depicted in the inset of (c). The inset is an overlay of the experimental image contrast.}
\end{figure*}

Absorption, PL and Raman spectra of the ternary Mo$_{1-x}$W$_{x}$S$_{2}$ monolayers are reported in Fig. \ref{fig:Raman}. Figure \ref{fig:Raman} a displays two strong peaks in the absorption spectra due to the A and B excitonic transitions, which arise from the spin-orbit-split bands at the K and K' points of the Brillouin zone. The energy difference of the A and B excitons originates mostly from the valence band splitting. While the B exciton transition energy increases monotonically with the tungsten content, the A exciton first slightly red- and then blue-shifts (see Fig. \ref{fig:Raman} b). Bowing of the exciton energies with $x$, being typical for ternary semiconductor alloys, has been previously observed in Mo$_{1-x}$W$_{x}$S$_{2}$ obtained by mechanical exfoliation of bulk crystals \cite{Chen2013} and by CVD \cite{Wang2016}. It is predicted also by first principle calculations.\cite{Xi2014} Fitting the exciton energies according to $E=(1-x) E_{MoS_2} + x E_{WS_2}- bx(1-x)$ yields slightly larger bowing parameters of $b$ = 0.25 $\pm$ 0.03 (A exciton) and $b$ = 0.37 $\pm$ 0.03 (B exciton) than previously reported. The bowing effect is also observed in the PL spectra (see Fig. \ref{fig:Raman} c). The lowest energy  of the emission peak which is associated to the A exciton is found for a ternary layer with $x$ = 0.23. A small peak due to B exciton emission is discernible  only in the spectrum of the binary MoS$_{2}$. The PL line width and the Stokes shift (ca. 20 meV) are comparable for all samples indicating that alloy broadening is not substantial. Both, the SOC strength and the PL emission energy are tunable via the composition. The spin-orbit splitting of the A and B exciton monotonically increases from 150 meV ($x=$ 0) to 425 meV ($x=$ 1). The emission energy of the Mo$_{1-x}$W$_{x}$S$_{2}$ is tunable from 1.88 eV ($x=$ 0.23) to 2.0 eV ($x=$ 1). 
Figure \ref{fig:Raman} d shows the corresponding Raman spectra of Mo$_{1-x}$W$_{x}$S$_{2}$ monolayers for $x$ = 0...1. The measurements were performed with a confocal Raman microscope (XploRA, Horiba Ltd.) and a 2400 lines/mm  grating with a calibration fidelity $<$ 3.4 cm$^{-1}$. The $A'_{1}$ mode associated to the out-of plane vibration of the sulfur atoms is found at 404.5 cm$^{-1}$ in MoS$_{2}$ and at 417.0 cm$^{-1}$ in WS$_{2}$. The $E'$ mode due to the in-plane vibration of the metal and sulfur atoms peaks at 384 cm$^{-1}$ in MoS$_{2}$ and at 353.5 cm$^{-1}$ in WS$_{2}$ in agreement with the reported values.\cite{Terrones2014} The $E'$ mode shows a clear two-mode behavior as also observed for CVD grown samples.\cite{Wang2016} Though the MoS$_{2}$-like and WS$_{2}$-like $A'_{1}$ modes approach each other, they also remain clearly split as visible by the presence of a shoulder in the $A'_{1}$ peaks of the monolayers with intermediate tungsten content $x$. The spectral weight of the Raman peaks changes according to the composition of the monolayer. The inset of Fig. \ref{fig:Raman} d presents a Raman map of the intensity ratio of the MoS$_{2}$-like and the WS$_{2}$-like $E'$ peak recorded for a sample with $x$ = 0.55. It demonstrates the homogeneity of the composition. The small fluctuations are due to the signal noise.

Next we discuss the structural and optical properties of TMDC heterostructures combining metallic TaS$_{2}$ with semiconducting MoS$_{2}$. TaS$_{2}$ can serve as semi-transparent electrode for the visible and near-infrared spectral region in all TMDC devices as shown in Fig. \ref{fig:Tantal} a. Presented is a transmission spectrum of a TaS$_{2}$ film of (2.97 $\pm$ 0.04) nm thickness with a sheet resistance of 2 k$\Omega$ which corresponds to a conductivity of 1.7 $\cdot$ 10$^5$ S/m. The film thickness was determined by X-ray reflectivity. The absorption spectrum of the MoS$_{2}$/TaS$_{2}$ heterostructure is a superposition of the absorption of MoS$_{2}$ and TaS$_{2}$ (see Fig. \ref{fig:Tantal} a). The excitonic absorption features of the A and B excitons of MoS$_{2}$ are clearly visible indicating that MoS$_{2}$ grows on TaS$_{2}$ in its semiconducting 2H phase.

More detailed information on the structural properties of the TaS$_{2}$/MoS$_{2}$ heterostructure are obtained by cross-sectional HRTEM combined with energy-dispersive X-ray spectroscopy (EDXS). Samples were prepared by conventional mechanical treatment and finally by Ar ion milling. HRTEM imaging was performed with a JEOL TEM/STEM 2200FS. At the same instrument STEM-based EDXS (Bruker) was conducted to visualize the distribution of the chemical elements. Elemental maps were recorded at a probe size of 1.5 nm for 360 s. Figure \ref{fig:Tantal} b shows the distribution of tanatalum (given in red) superimposed to the distribution of sulfur and molybdenum (given in blue). Mo and S cannot be disentangled since the energy of the K ionization edge of S (2.3 keV) exactly matches the energy of the L ionization edge of Mo. The Ta signal only shows up in the lower part of the about 10 nm thick film. The upper part comprises both, Mo and S. This hints at well-separated phases of TaS$_{2}$ (lower part) and MoS$_{2}$ (upper part). More evidence is gained from the HRTEM images of Fig. \ref{fig:Tantal} d and e recorded at two different sample positions. An atomically sharp interface between TaS$_{2}$ and MoS$_{2}$ is clearly visible by the transition from the darker to the brighter contrast. The first layers of TaS$_{2}$ grow conformally on the fused silica substrate which shows some waviness. The lattice distortions mostly vanish from the fifth layer onwards (see Fig. \ref{fig:Tantal} d). In MoS$_{2}$, the formation of stacking faults is observed (see left part in Fig. \ref{fig:Tantal} d). They indicate the location of grain boundaries. The amorphous substrate does not induce preferred in-plane orientations of the grains. Therefore, the crystal structure of the individual (mono)layers becomes visible only at sample regions where the direction of the electron beam incidentally coincides with a zone axis of the TMDC lattice parallel to the substrate. An example is given in Fig. \ref{fig:Tantal} e. Here, the crystalline structure of the individual MoS$_{2}$ layers is clearly visible while it is blurred in the TaS$_{2}$ layers indicating different in-plane orientations. The weak van der Waals interaction does apparently not suffice to induce an epitaxial relationship between the MoS$_{2}$ and TaS$_{2}$ layers and therefore the crystal lattices are twisted with respect to each other as schematically depicted in Fig. \ref{fig:Tantal} c. Respective image contrast simulations (see Fig. \ref{fig:Tantal} f) assuming a twist angle of 20$^\circ$ between the lattices and with the electron beam parallel to the MoS$_{2}$ $\left[ 1\overline{2}10 \right]$ zone axis reproduce the HRTEM image in Fig. \ref{fig:Tantal} e very well. Details on the simulations are given in the supportive information.  

In conclusion, we have presented a versatile method to grow binary and ternary monolayers as well as heterostructures of different TMDCs. The method is based on pulsed direct resistive heating and sublimation of the metal elements. The amount of the deposited material as well as the composition is precisely controllable by the pulse length and pulse sequence. The current metal evaporators show reproducible characteristics for more than 300 operation cycles, thus fulfilling stability requirements for the MBE sources. The TMDCs grow in a layer-by-layer fashion resulting in homogeneous films, in contrast to CVD yielding discontinuous films composed of mono- and multilayer flakes. Band gap and SOC engineering is demonstrated by alloying of MoS$_{2}$ and WS$_{2}$ over the entire composition range. Heterostructures composed of TaS$_{2}$ and MoS$_{2}$ exhibit atomically abrupt interfaces. Since the TMDCs are grown by sublimation of the constituent elements, complex alloys and heterostructures can thus be prepared in a single process paving the way to all-TMDC based optoelectronic, photonic and valleytronic devices.

\begin{acknowledgments}
This work was funded by the Deutsche Forschungsgemeinschaft (DFG, German Research Foundation) - Projektnummer 182087777 - SFB 951.
\end{acknowledgments}

\end{document}